# Dramatic Raman Gain Suppression in the Vicinity of the Zero Dispersion Point in Gas-Filled Hollow-Core Photonic Crystal Fiber

S. T. Bauerschmidt[1], D. Novoa[1], and P. St.J. Russell[1,2]

[1]Max Planck Institute for the Science of Light, Guenther-Scharowsky-Str. 1, 91058 Erlangen, Germany
[2]Department of Physics, University of Erlangen-Nuremberg, Germany

In 1964 Bloembergen and Shen predicted that Raman gain could be suppressed if the rates of phonon creation and annihilation (by inelastic scattering) exactly balance. This is only possible if the momentum required for each process is identical, i.e., phonon coherence waves created by pump-to-Stokes scattering are identical to those annihilated in pump-to-anti-Stokes scattering. In bulk gas cells, this can only be achieved over limited interaction lengths at an oblique angle to the pump axis. Here we report a simple system that provides dramatic Raman gain suppression over long collinear path-lengths in hydrogen. It consists of a gas-filled hollow-core photonic crystal fiber whose zero dispersion point is pressure-adjusted to lie close to the pump laser wavelength. At a certain precise pressure, generation of Stokes light in the fundamental mode is completely suppressed, allowing other much weaker nonlinear processes to be explored.

Hollow-core photonic crystal fibers (PCFs) [1, 2] have in recent years emerged as an ideal vehicle for enhancing gas-based nonlinear optics. This is because they offer tight modal confinement, long well-controlled interaction lengths, a very high damage threshold and precise pressure control of the dispersion and nonlinearity [3, 4]. Broad-band guiding kagomé-style PCF, in particular, has been used for extreme pulse compression, generation of tunable deep and vacuum UV light, and ultra-broadband supercontinuum generation. When filled with Raman-active gases and vapors, it has also been used for efficient wavelength conversion [4-6], Cs-based quantum memories [7] and frequency comb [8-10] and supercontinuum [11] generation.

It has recently been demonstrated that the unique S-shaped dispersion curve, in the vicinity of the pressure-tunable zero dispersion point (ZDP) in gas-filled kagomé-PCF, permits optical phonons created by pump-to-Stokes scattering (from $\omega_{p1}$ to $\omega_{p1} - \Omega_R$, where $\Omega_R/2\pi$ is the Raman frequency) to perfectly phase-match anti-Stokes scattering at a different pump frequency, i.e., from $\omega_{p2}$ to $\omega_{p2} + \Omega_R$. This was used to achieve efficient broad-band frequency up-conversion of a signal at $\omega_{p2}$ [6].

Raman gain suppression has previously been studied theoretically [12, 13] and observed in free space configurations as a dark ring in the emission cone of the anti-Stokes signal [14]. In this Letter we describe how at a certain pressure, which depends on the fiber structure, the gas dispersion and the pump wavelength, it can be arranged that $\omega_{p1} = \omega_{p2}$, i.e., the optical phonon created in pump-to-Stokes conversion is identical to the one annihilated in pump-to-anti-Stokes conversion. The resulting precise balancing of phonon creation and annihilation results in perfect suppression of the Raman gain. If this condition is not perfectly fulfilled, as is normally the case in stimulated Raman scattering (SRS), the Stokes signal is amplified above threshold according to $I_{s0} \exp(\gamma_{eff} I_p z)$, where $\gamma_{eff}$ is the effective Raman gain, $z$ the propagation distance, $I_p$ the pump intensity and $I_{s0}$ the initial value of the Stokes intensity. This is accompanied by the excitation of a population of optical phonons in the form of a "coherence wave" of synchronous molecular oscillations with a well-defined wavelength that is set by the optical dispersion.

The effective gain may be written in the form:

$$\gamma_{eff} = \rho S_{ij} g_P \qquad (1)$$

where $S_{ij}$ is the spatial overlap integral between the $LP_{01}$ pump and the $LP_{ij}$ Stokes mode, $g_P$ is the Raman gain of the gas and $\rho$ is the gain reduction factor, which for no pump depletion and negligible Kerr nonlinearity can be expressed in the form [15, 16]:

$$\rho = \left| \mathrm{Re} \sqrt{\left(\frac{p-1}{2}\right)^2 - \left(\frac{\vartheta_{01}}{g_P I_P}\right)^2 - i(p+1)\frac{\vartheta_{01}}{g_P I_P}} - \left(\frac{p-1}{2}\right) \right| \quad (2)$$

where $p = g_{AS}\omega_{AS}/(g_P\omega_P) > 1$ and $g_i$ is the Raman gain at pump frequency $\omega_i$. Gain suppression (i.e. $\rho = 0$) occurs when the dephasing parameter $\vartheta_{01} = \beta_{01}(\omega_{AS}) + \beta_{01}(\omega_S) - 2\beta_{01}(\omega_P)$ is zero, where $\beta_{ij}(\omega)$ is the propagation constant of the $LP_{ij}$ mode at frequency $\omega$. In a kagomé-PCF, $\beta_{ij}(\omega)$ can be approximated to good accuracy by:

$$\beta_{ij} = \sqrt{k_0^2 n_{gas}^2(p,\lambda) - u_{ij}^2/a^2(\lambda)} \qquad (3)$$

where $k_0 = 2\pi/\lambda$ is the vacuum wavevector, $n_{gas}$ the refractive index of the filling gas, $p$ the gas pressure, $u_{ij}$ the $j$th root of the $i$th order Bessel function of the first kind and $a(\lambda) = a_{AP}(1+t\lambda^2/(a_{AP}d))$, where $a_{AP}$ is the area-preserving core radius, $t = 0.06$ is an empirical parameter derived from finite element modeling of an idealized kagomé structure and $d$ is the core wall thickness [17].

Therefore $\vartheta_{01}$ depends directly on gas pressure [6], allowing precise tuning of the gain suppression point without need for the complicated crossed-beam arrangements used in free space configurations. Although Raman gain suppression has been studied in solid core fibers, it is severely impaired by the broad-band nature of the Raman gain and by strong four-wave mixing [18, 19].

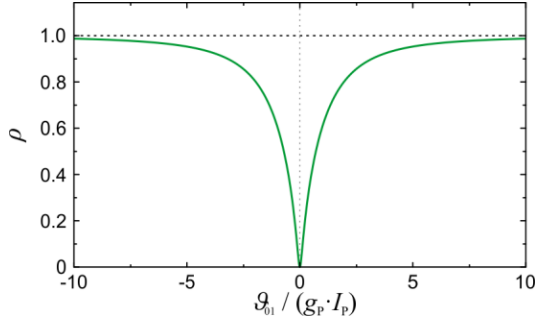

FIG. 1 The gain reduction factor $\rho$ is plotted against the normalized dephasing parameter $(\vartheta_{01}/g_P I_P)$.

The gain reduction factor $\rho$ is plotted in Fig. 1 against $(\vartheta_{01}/g_P I_P)$ for $p=1.87$, measured experimentally via the Raman gain [20]. Note that $S_{01}=1$ since all the signals are in the $LP_{01}$ mode.

The $LP_{01}$ dispersion curve for perfect gain suppression is shown in Fig. 2, for a kagomé-PCF with 30 μm core diameter filled with hydrogen at 24.7 bar. The Raman frequency shift is $\Omega_R/2\pi \approx 125$ THz, which corresponds to the fundamental vibrational mode of hydrogen. Perfect gain suppression is expected for a pump signal at 683 nm, which is very close to the ZDP at 698 nm. Under these conditions the phonon four-vectors (indicated by green dashed lines) for pump-to-Stokes and pump-to-anti-Stokes scattering in the $LP_{01}$ mode are identical.

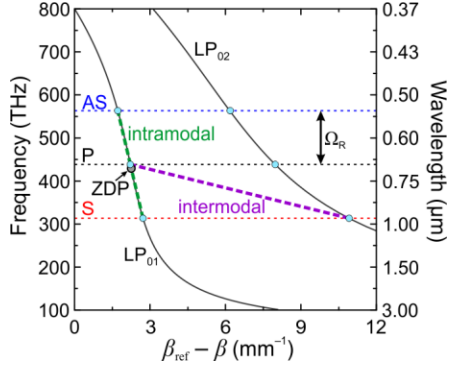

FIG. 2 Dispersion diagram for a kagomé-PCF at a hydrogen-filling pressure of 24.7 bar. Dispersion curves for $LP_{01}$ and $LP_{02}$ mode are shown. In order to magnify the small, but crucial deviation from an ideal linear dispersion relation, we plot against $(\beta_{ref}-\beta)$, where $\beta_{ref}$ is a linear function of frequency chosen such that $(\beta_{ref}-\beta_{01})$ is zero at 800 THz.

To demonstrate gain suppression experimentally, a 90 cm long hydrogen-filled kagomé-PCF with 30 μm core diameter was pumped with 1.6 ns, 2.5 μJ laser pulses at 683 nm (Fig. 3). These pulses were generated via SRS from a 532 nm laser in a second length of hydrogen-filled kagomé-PCF. The output spectrum was monitored with an optical spectrum analyzer and the near-field profiles of the modes imaged at the fiber endface using a CCD camera (see Fig. 4a&b).

The photon flux (i.e., the number per second) in the three signals at the fiber output was recorded while scanning the hydrogen-filling pressure (see Fig. 4b). Although the Stokes signal at 955 nm dropped by about 10 dB in the vicinity of the calculated gain suppression pressure, it did not completely disappear as expected. Strong suppression was however clearly apparent in the anti-Stokes signal at 532 nm, which vanished at this pressure. It turned out that the Stokes signal exactly at the gain suppression point was in the $LP_{02}$ mode, the $LP_{01}$ mode having been completely suppressed. Tuning the pressure away from this point, the Stokes light underwent a smooth transition from the $LP_{02}$ to the $LP_{01}$ mode (see Fig. 4a). Stokes emission into the $LP_{02}$ mode occurs via the intermodal coherence wave marked in Fig. 2 by the purple dashed line. It turns out that the intermodal overlap (and hence the gain) is larger for the $LP_{02}$ mode ($S_{02}=0.85$) than for any other higher order mode. As a result, when the $LP_{01}$ gain is suppressed, the $LP_{02}$ mode is the first to appear because its net gain (taking account of gain and loss) is the highest of any of the higher order modes (see gray-shaded region in Fig. 4b).

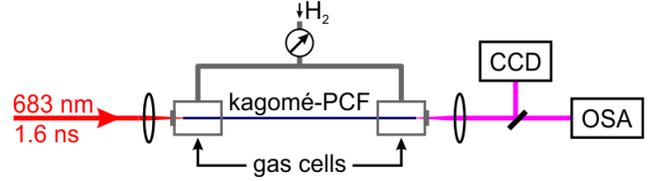

FIG. 3 Sketch of the experimental setup. OSA: Optical spectrum analyzer; CCD: camera.

To model the system in the gain suppression region, we used a semi-classical treatment of SRS [21] based on a multimode-extended set of coupled Maxwell-Bloch equations, namely

$$\frac{\partial}{\partial z}E_{\sigma,l} = -\kappa_{2,l}\frac{\omega_l}{\omega_{l-1}}\sum_{\nu\xi\eta}^{M} i\frac{s_{\sigma\nu\xi\eta}}{s_{\nu\xi}}Q_{\nu\xi}E_{\eta,l-1}q_{\eta,l-1}q_{\sigma,l}^* \\ -\kappa_{2,l+1}\sum_{\nu\xi\eta}^{M} i\frac{s_{\sigma\nu\xi\eta}}{s_{\nu\xi}}Q_{\nu\xi}^*E_{\eta,l+1}q_{\eta,l+1}q_{\sigma,l}^* - \frac{1}{2}\alpha_{\sigma,l}E_{\sigma,l} \quad (4)$$

$$\frac{\partial}{\partial \tau}Q_{\nu\xi} = -Q_{\nu\xi}/T_2 + in\frac{1}{4}s_{\nu\xi}\sum_l \kappa_{1,l}E_{\nu,l}E_{\xi,l-1}^*q_{\xi,l}q_{\nu,l-1}^* \quad (5)$$

where the integer $l$ denotes the sideband at frequency $\omega_l = \omega_P + l\,\Omega$ and the summations go over all possible permutations of the modal set $M$. We describe the complex electric field amplitude $e_{\sigma,l}(x,y,z,\tau) = F_\sigma(x,y)E_{\sigma,l}(z,\tau)q_{\sigma,l}$ of a guided mode (mode index $\sigma$) in terms of its normalized transverse spatial profile $F_\sigma(x,y)$, the slowly varying field envelope $E_{\sigma,l}(z,\tau)$ and the fast oscillating phase term $q_{\sigma,l} = \exp[-i\beta_\sigma(\omega_l)z]$. $\tau$ is the time relative to a frame travelling with the group velocity of the pump pulse. The coupling constants $\kappa_{1,l} = c\varepsilon_0(2g_l/NT_2\hbar\omega_{l-1})^{0.5}$ and $\kappa_{2,l} = \kappa_{1,l}N\hbar\omega_{l-1}/2c\varepsilon_0$ can be derived from experimentally measured gain values $g_l$ [20] where $N$ is the molecular number density, $c$ the vacuum speed of light, $\hbar$ the reduced Planck's constant, $\varepsilon_0$ the vacuum permittivity and $T_2$ is the dephasing time of the Raman coherence [20]. Since the highest coherence achieved in the experiment is of order 0.001, the population imbalance $n$ is assumed to be −1, i.e., the majority of hydrogen molecules remain in the ground state.

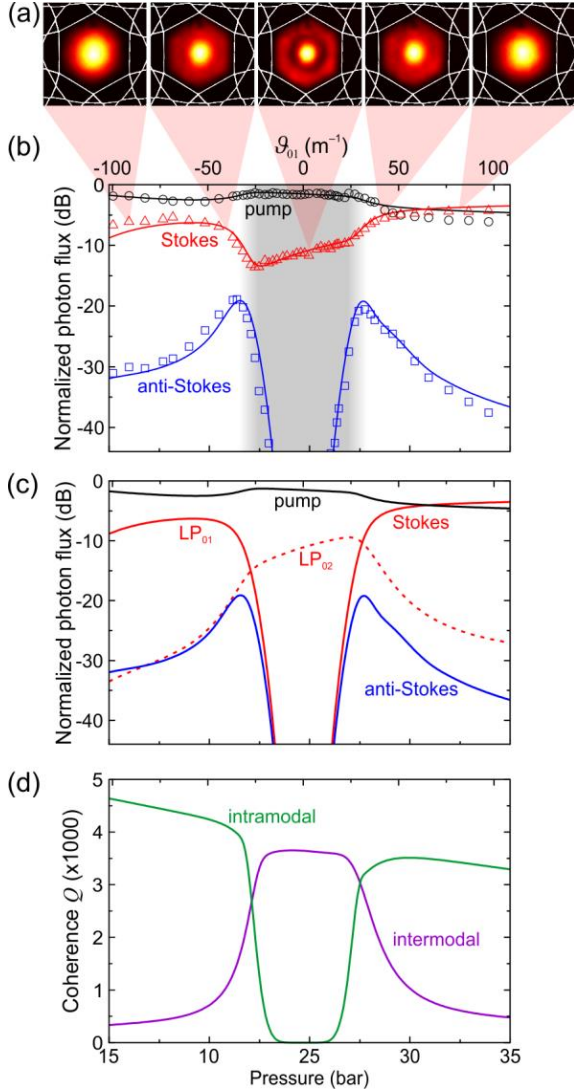

FIG. 4 (a) Near-field profiles of the Stokes light at the fiber endface recorded at the different pressures and overlaid onto a scanning electron micrograph of the fiber microstructure. (b) Measured (symbols) and simulated (solid lines) photon flux (number per second) in the pump, Stokes, and anti-Stokes bands, plotted against increasing pressure and normalized to the launched flux of pump photons. Stokes light is emitted in the $LP_{02}$ mode within the gray-shaded region. (c) Simulated photon flux of pump, Stokes, and anti-Stokes bands in the $LP_{01}$ (solid lines) and $LP_{02}$ (dashed line) mode plotted against increasing pressure and normalized to the launched pump signal. (d) Simulated maximum value of intramodal and intermodal coherence $Q$ plotted against increasing pressure.

The generalized nonlinear spatial overlap integrals are given by [22]

$$s_{\sigma\nu\xi\eta} = \frac{\int F_\sigma^* F_\nu^* F_\xi F_\eta dA}{\left(\int |F_\sigma|^2 dA \int |F_\nu|^2 dA \int |F_\xi|^2 dA \int |F_\eta|^2 dA\right)^{1/2}} A_{\text{eff}} \quad (6)$$

and

$$s_{\nu\xi} = \frac{\int |F_\nu|^2 |F_\xi|^2 dA}{\left(\int |F_\nu||F_\xi| dA\right)^2} A_{\text{eff}}, \quad (7)$$

where $A_{\text{eff}}$ is the effective mode area of the pump in the fundamental mode [23] and the integrals are evaluated over the transverse cross-section of the fiber. The transverse mode field profiles $F_\sigma(x,y)$ are obtained via finite element modeling of an idealized kagomé structure and are assumed to be frequency independent in the spectral range considered in this work.

The exponential loss rate (per m) of each mode and sideband is represented by $\alpha_{\sigma,l}$. For the fundamental mode these values were experimentally determined using a standard cut-back method, yielding values of order a few dB/m.

The first Stokes signal is seeded mainly by scattering from thermally excited phonons. For simplicity in the simulations we assume a uniform noise-floor of $E_{\text{noise}} = 50$ V/m for all the Raman lines and modes except the fundamental pump mode, a simplification that was successfully employed in a previous study [6].

With this, the spatio-temporal dynamics of the optical fields and the Raman coherence $Q$ were modeled, and for simplicity only the $LP_{01}$ and $LP_{02}$ modes were included in the analysis. Nevertheless the results are in remarkable agreement with experimental data (see Fig. 4b), the only free parameter being the attenuation of the $LP_{02}$ mode at the Stokes frequency, which was taken to be 17 dB/m, a value in agreement with cut-back measurements of the $LP_{02}$ mode loss for a similar kagomé-PCF [24].

The good agreement between theory and experiment allows the dynamics of the $LP_{01}$ and $LP_{02}$ Stokes modes to be investigated separately (Fig. 4c), as well the intra- and intermodal coherence $Q$ (Fig. 4d). The results show that the $LP_{01}$ Stokes and anti-Stokes signals drop dramatically close to the gain suppression point (red solid line in Fig. 4c), while at the same time the $LP_{02}$ signal (red dashed line in Fig. 4c) is enhanced. At large values of dephasing, in contrast, the $LP_{02}$ signal is ~20 dB weaker than the $LP_{01}$ signal, due to a lower spatial overlap and higher loss. Approaching the gain suppression point, the $LP_{01}$-$LP_{01}$ intramodal coherence drops strongly (Fig. 4d) while the $LP_{01}$-$LP_{02}$ intermodal coherence becomes dominant.

As predicted in 1964 by Bloembergen and Shen [12] but never observed before in a collinear arrangement, the $LP_{01}$ anti-Stokes signal peaks at two values of pressure above and below the gain suppression point. This is because the coherent phonons created by pump-to-Stokes conversion can immediately be used for pump-to-anti-Stokes conversion, even though slightly dephased. Nevertheless, conversion to the anti-Stokes frequency can be reasonably efficient (~ 1% in the experiment).

The conversion efficiency to the $LP_{02}$ Stokes band peaked at 10% at a launched pump energy of 2.5 μJ and a pressure of 24.5 bar (see Fig. 5a), in good agreement with the numerical simulations. This peak arises because at high energy the Stokes light is generated close to the fiber input and is depleted by high LP$_{02}$ loss, whereas at low energy the Raman gain is weak. As a result, for a given fiber length there is a pulse energy at which the conversion efficiency is maximum.

This fall-off in conversion efficiency at high pump energies was already observed in the first demonstration of SRS in hydrogen-filled hollow-core PCF, although it was attributed to Raman-enhanced self-focusing [2]. Note that this phenomenon can be seen even if the pump wavelength and filling-pressure do not fulfill the conditions for perfect gain-suppression; indeed, both frequency and pressure ranges of gain-suppression widen linearly with pump power. The pressures and optical frequencies for perfect gain suppression are plotted in Fig. 5b for core diameters between 20 and 60 µm

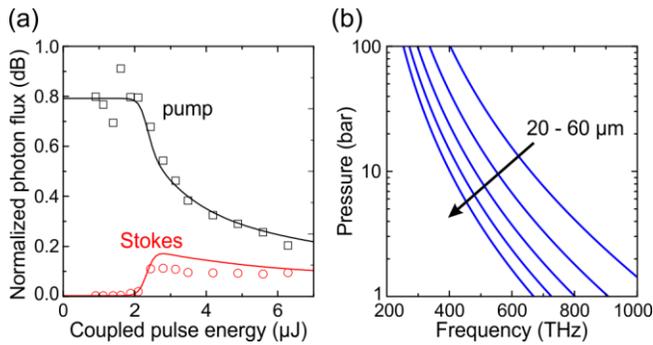

FIG. 5 (a) Pump and Stokes photon fluxes (normalized to the launched pump photon flux) plotted against pump energy at 24.5 bar. The measured data (squares and circles) are compared to the numerical results (solid lines). (b) Pressure-frequency dependence of the point of perfect gain suppression for five different core diameters between 20 and 60 µm. Different curves correspond to fibers with different hollow-core diameters.

In conclusion, gas-filled hollow-core kagomé-PCF provides a collinear system for observing the dramatic suppression of Raman gain first predicted by Bloembergen and Shen. This is achieved by operating in the vicinity of the pressure-tunable zero dispersion point, where at a specific pressure and pump frequency the phonon coherence wave for pump-to-Stokes conversion exactly matches that for pump-to-anti-Stokes conversion. Suppression of Raman gain in a collinear system provides a route for investigating spontaneous Raman scattering at high pump energies, a regime that has so far been inaccessible because of the inevitable onset of SRS. Such studies may have important implications in quantum information, and lead to a deeper understanding of Raman scattering as a noise source in quantum systems. Strong suppression of LP$_{01}$ Stokes amplification also allows clean Raman amplification of higher order fiber modes [23, 25], which could open new avenues for particle trapping and manipulation in hollow-core PCF [26, 27]. Finally, by tuning away from the point of perfect gain suppression, strong anti-Stokes bands can be generated, as predicted in 1964, but until now never exploited.